\documentclass[prl,a4paper,twocolumn,superscriptaddress]{revtex4}
\usepackage{graphicx}
\input{epsf}

\begin{document}

\title{Microscopic Superfluidity in Bose Gases: From 3D to 1D} 

\author{Kwangsik Nho}
\affiliation{Department of Physics and Astronomy, Washington State University,
         Pullman, Washington 99164-2814, USA}
\author{D. Blume}
\affiliation{Department of Physics and Astronomy, Washington State University,
         Pullman, Washington 99164-2814, USA}

\begin{abstract}
The superfluid fraction of ideal and interacting inhomogeneous Bose gases 
with varying asymmetry is investigated 
at finite temperature using well-known properties of the harmonic
oscillator as well as the 
essentially exact microscopic path integral Monte Carlo method. 
We find that
the superfluid fraction (i)
is essentially independent of the interaction
strength for all temperatures considered,
(ii) 
changes profoundly as the effective
dimensionality is varied from three- to 
one-dimensional, 
(iii) 
is approximately
equal to the condensate fraction $N_0/N$ for spherical Bose gases, 
and
(iv) 
deviates dramatically from $N_0/N$
for highly-elongated Bose gases.
\end{abstract}

\maketitle

Macroscopic objects such as liquid $^4$He show many peculiar
properties that can be attributed to superfluidity~\cite{donn95}.
Among these are the absence of viscosity, the occurence of 
persistent currents, the existence of vortices and the reduction of
the moment of inertia.
Connections between manifestations of superfluidity and
Bose Einstein condensation have been studied extensively in the
context of liquid $^4$He since the discovery of its superfluidity in 1938.
While much progress has been made in our understanding
of such strongly interacting systems, many questions remain unanswered.

Thanks to the realization of gaseous Bose Einstein condensates in
1995~\cite{ande95}, 
the study of superfluid effects of mesoscopic systems has become
possible. Indeed, the creation of vortices~\cite{matt99} 
and vortex lattices~\cite{abos01} 
has been demonstrated in inhomogeneous Bose gases, and, most recently, also
in degenerate Fermi gases in the BEC-BCS crossover regime~\cite{zwie05}. 
Following the work on $^4$He enclosed in a cylinder~\cite{baym69}, 
superfluidity of inhomogeneous systems can be described through their 
rotational properties. The superfluid fraction is defined by the 
departure of the quantum mechanical moment of inertia $\Theta_{\hat{n}}$ 
with respect to $\hat{n}$
from
its classical, or rigid, value $\Theta^{rig}_{\hat{n}}$.
Here, the moment of inertia $\Theta_{\hat{n}}$,
$\Theta_{\hat{n}} = (  \partial \langle \vec{L} \cdot \hat{n} \rangle_{\omega}/
\partial \omega)_{\omega=0}$,
is defined by the linear
response of the system to a rotational field  
$H_{ext}=-\vec{\omega} \cdot \vec{L}$,
where
$\vec{\omega}=\omega \hat{n}$;
$\omega$ denotes the angular frequency and $\vec{L}$ the 
total angular momentum.
The thermal expectation value $\langle \cdot \rangle_{\omega}$ 
is
evaluated for the system perturbed by $H_{ext}$.
The normal fraction is the part of the system that responds 
classically, i.e., $\Theta_{\hat{n}} / \Theta^{rig}_{\hat{n}}$,
and the superfluid fraction 
$(n_s/n)_{\hat{n}}$ is $1-\Theta_{\hat{n}} / \Theta^{rig}_{\hat{n}}$.

The temperature dependence of 
$\Theta_{\hat{n}}$
has been evaluated 
for a non-interacting Bose gas under 
harmonic confinement in the so-called macroscopic approximation~\cite{stri96}.
The effects of weak interactions have been estimated 
within the Thomas-Fermi
approximation~\cite{stri96}. 
This Letter determines the temperature-dependence of 
$\Theta_{\hat{n}}$
and $\Theta^{rig}_{\hat{n}}$, 
and hence of the superfluid fraction, for small 
atomic gases with $N=27$ bosons for varying confinement
and interaction strength non-perturbatively
using the essentially exact microscopic path
integral Monte Carlo (PIMC) method~\cite{cepe95}. 
For the ideal gas, we additionally determine thermal 
expectation values using well-known properties of the harmonic 
oscillator~\cite{groo50}.
In contrast to $^4$He clusters~\cite{sind89} 
or deformed nuclei~\cite{bohr75}, whose  
interaction strength
and internal temperature are largely ``set by nature'',
atomic gases provide us with unprecedented control.
The temperature can be controlled by changing the
cooling scheme~\cite{gerb04},
the interaction strength can be tuned by applying an
external magnetic field in the vicinity of a Feshbach
resonance~\cite{inou98,corn00}, 
and the dimensionality can be reduced by varying the
external confinement~\cite{goer01,grei01}. 
Here, we focus on the crossover from three-dimensional (3D) to
one-dimensional (1D) behavior. We show that 
reduced dimensionality leads to an increase of the superfluid response.
The superfluid fraction for 3D gases roughly coincides with the
condensate fraction $N_0/N$.
In the quasi-1D regime, however,
the superfluid fraction is much larger than $N_0/N$.
Our calculations show that
the superfluid response depends, if at all, weakly on the
strength of the atom-atom
interactions.

Consider $N$ bosons with mass $m$ under external  
harmonic confinement,
\begin{eqnarray}
H =  \sum_{j=1}^N \left[ \frac{-\hbar^2}{2m} \nabla_j^2
+ \frac{1}{2}m (
 \omega_{\rho}^2 \rho_j^2
+\omega_z^2 z_j^2 
) \right]
+ \sum_{j<k}^N V(r_{jk}).
\end{eqnarray}
Here, $\rho_j$ and $z_j$ denote the transverse and longitudinal coordinate
of the $j$th atom, respectively, and 
$\omega_{\rho}$ and $\omega_z$ the 
transverse and longitudinal 
angular frequency of the trapping potential, respectively.
The atom-atom potential $V$ depends on the interparticle
distance $r_{jk}$ between atom $j$ and atom $k$. 
For the non-interacting gas, i.e., $V(r)=0$, we
calculate thermal expectation values in the grandcanonical ensemble
using well-known properties of the harmonic 
oscillator~\cite{groo50}.
To simulate effectively repulsive Bose gases, we use a hard sphere 
potential $V(r)$ with 3D atom-atom 
scattering length $a$; in particular,
$a=0.00433$ and $0.0433a_z$, where $a_z=\sqrt{\hbar/(m \omega_z)}$.
In this case, we use
the numerically more involved PIMC technique~\cite{cepe95},
which determines thermal expectation values in the canonical
ensemble.
For the purpose of the present study, differences between
expectation values calculated in the
grandcanonical and in the canonical 
ensemble (see also Ref.~\cite{herz97}) are negligible.

\begin{figure}[tbp]
\vspace*{-1.75in}
\rotatebox{0}{\centerline{\epsfxsize=3.25in\epsfbox{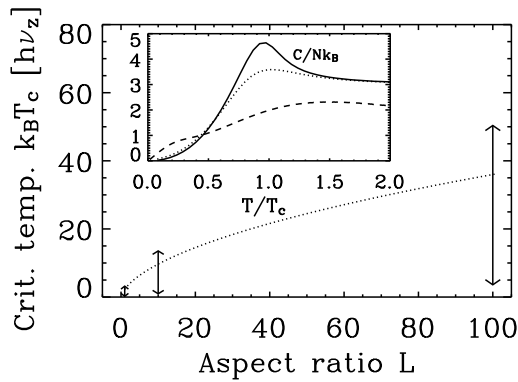}}}
\vspace*{-.175in}
\caption{Approximate 3D transition 
temperature $k_B T_c$ (see, e.g., Eq.~(19) of Ref.~\protect\cite{dalf99})
in units of $\hbar \omega_z$ ($\omega_z=2 \pi \, \nu_z$) as a function
of $L$ for $N=27$.
Vertical arrows indicate the 
interval $0.1 \le T/T_c \le 1.4 $
for
$L=1$, 10 and 100.
The inset shows the specific heat $C$ divided by 
$N k_B$ calculated in the grandcanonical ensemble
as a function of $T/T_c$
for $N=27$ and $L=1$ (solid line), 10 (dotted line) and 100 (dashed line).
}
\label{fig1}
\end{figure}

To investigate the crossover from 3D to 1D 
for $N$ bosons, we
vary the angular frequency $\omega_{\rho}$ 
such that $L=1$,
$10$ and $100$, where $L = \omega_{\rho}/\omega_z$.
The approximate 3D transition temperature $T_c$, obtained for
vanishing atom-atom interactions, then depends on $\omega_{\rho}$, $\omega_z$ 
and $N$ (the $T_c$ used throughout this
paper includes finite-size corrections; see, e.g., 
Eq.~(19) of Ref.~\cite{dalf99}).
A dotted line in Fig.~\ref{fig1} shows $T_c$ as a function of 
the aspect ratio $L$ for $N=27$.
Below, we report calculations for $L=1$, 10 and 100 over a wide temperature
range, i.e.,
$0.1 
\lesssim 
T/T_c \lesssim 1.4$ (see vertical arrows in Fig.~\ref{fig1}).

Highly-elongated gases at $T=0$ can to a very good
approximation be described by an effective
1D Hamiltonian for any 3D scattering length $a$ if
$N/L \ll 1$~\cite{petr04}. For $N=27$ and $L=100$, we find $N/L=0.27$.
At finite temperature, the behavior of highly-elongated Bose gases
depends on two energy scales, the oscillator energies
$\hbar \omega_{\rho}$ and $\hbar \omega_z$ of the tight and
weak confinement direction, respectively.
For $N=27$ and $L=100$, 
three temperature regimes exist~\cite{drut97}:
i) $T$ is larger than the 3D transition temperature $T_c$ 
(excited transverse modes are occupied); 
ii) $T$ is lower than $T_c$ but larger than the 1D transition temperature
$T_c^{1D}$~\cite{kett96} (transverse excitations are
largely frozen out); and
iii) $T$ is smaller than $T_{c}^{1D}$ 
(excited longitudinal modes are largely frozen out).
For $N=27$ and $L=100$,
the approximate 3D transition temperature is 
$k_B T_c=36.0 \hbar \omega_z$, while
the approximate 1D
transition temperature 
is $k_B T_c^{1D} = 9.67 \hbar \omega_z$, corresponding to $0.269 T_c$.

To understand the significance of the 3D transition temperature $T_c$
we calculate the specific heat $C$,
$C= (\partial U/\partial T)_N$, where
$U$ denotes the internal energy~\cite{drut97},
for the ideal gas in the grandcanonical ensemble.
The inset of Fig.~\ref{fig1} shows the specific heat $C$
for $N=27$ for three different aspect ratios, i.e.,
$L=1$ (solid line), 10 (dotted line)
and $100$ (dashed line).
Since the specific heat shows a 
peak, although broadened due to the finite size
of the Bose gas, at $T/T_c \approx 1$ for $L=1$ and 10,
and at $T/T_c \approx 1.4$ for $L=100$,
it is justified to speak of a 3D transition
temperature
for Bose gases with as few as $N=27$ atoms.
In contrast, 
the transition to 
macroscopic occupation of the
lowest energy state for the quasi-1D gas with $L=100$, 
i.e., to ``1D condensation'', does not imprint
a clear signature on the specific heat (see also Ref.~\cite{kett96,drut97}). 

\begin{figure}[h]
\vspace*{-0.71in}
\hspace*{.5in}
\rotatebox{0}{\centerline{\epsfxsize=5.75in\epsfbox{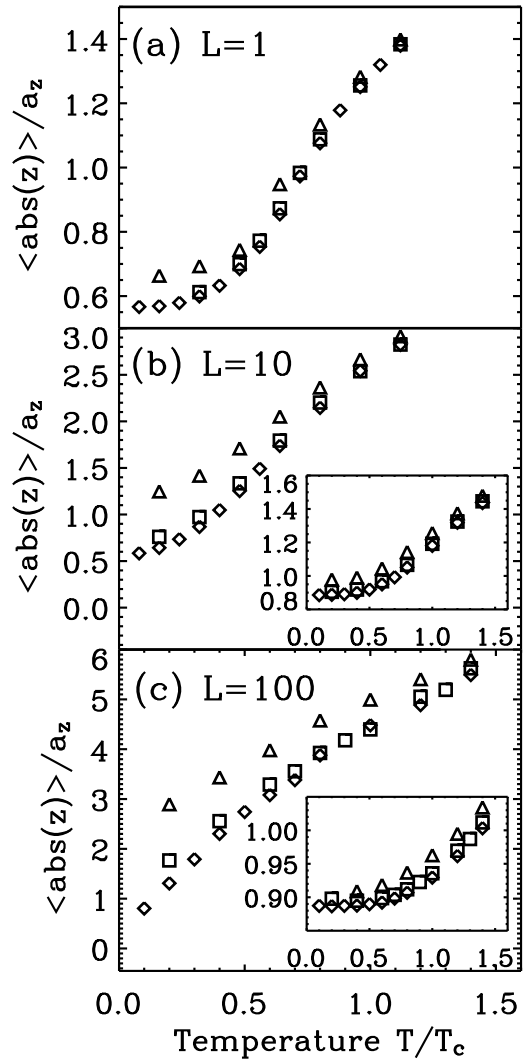}}}
\vspace*{-0.51in}
\caption{PIMC
expectation value of $|z|$ in units of
$a_z$ as a function of $T/T_c$ 
for $N=27$
for (a) $L=1$,
(b) $10$ and (c) $100$.
Diamonds show the results for $a/a_z=0$, 
squares those for $a/a_z=0.00433$
and triangles those for $a/a_z=0.0433$.
For $L=10$ and $L=100$, 
the insets show the expectation value of
$\rho$ in units of $a_{\rho}$ as a function of $T/T_c$.
Statistical uncertainties are smaller than the symbol
size~\protect\cite{remarkerr1}.}
\label{fig2}
\end{figure}

Figure~\ref{fig2} shows 
the expectation value of 
$|z|$ in units of $a_z$, calculated using the PIMC method, as a function of 
the scaled temperature $T/T_c$ for (a) $L=1$,
(b) $L=10$ and 
(c) $L=100$ for three scattering lengths;
$a=0$ (diamonds), $a=0.00433a_z$ (squares) and $a=0.0433a_z$ (triangles).
At low $T/T_c$, our expectation values of $|z|$
for $a=0$ (circles) approach the zero temperature value, i.e.,
$\langle |z| \rangle = 0.564 a_z$. 
Since the energy of the transverse excitations increases with
increasing $L$, 
the expectation value of $|z|$ for $L=100$ approaches the zero-temperature value
at a lower scaled temperature $T/T_c$ than that for $L=1$.
For repulsive interactions, 
i.e., $a/a_z=0.00433$ (squares) and
$a/a_z=0.0433$ (triangles), 
the expectation value of $|z|$ increases compared to that of
the non-interacting gas. 

The insets 
of Figs.~\ref{fig2}(b) and (c)
show the expectation value
of $\rho$ in units of $a_{\rho}$, where 
$a_{\rho}=\sqrt{\hbar/(m \omega_{\rho})}$,
for $L=10$ and $100$, respectively, as a function of 
$T/T_c$ (using the same symbols as in the main figure).
At $T=0$, the expectation value of $\rho$ is $0.886 a_{\rho}$ for
the non-interacting gas.
Just as the expectation value of $|z|$, that
of $\rho$ depends strongly on the interaction strength $a$. 
The nearly constant expectation value of $\rho$ for $L=100$ (for
a given value of $a$)
at low 
$T/T_c$
indicates that
the excitations in the transverse direction are frozen out
for $T \lesssim 
0.4 T_c$, and hence for $T < T_c^{1D}$.

\begin{figure}
\vspace*{-0.71in}
\hspace*{.5in}
\rotatebox{0}{\centerline{\epsfxsize=5.75in\epsfbox{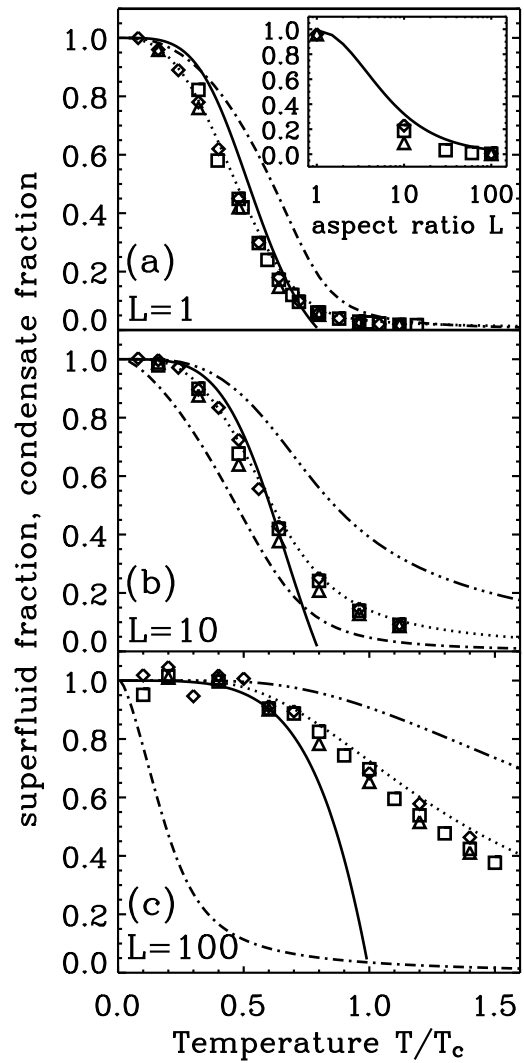}}}
\vspace*{-.51in}
\caption{
Superfluid fraction $(n_s/n)_{\hat{z}}$
for $N=27$ and (a) $L=1$,
(b) $L=10$ and (c) $L=100$:
Diamonds show the PIMC results for $a/a_z=0$, 
squares those for $a/a_z=0.00433$
and triangles those for $a/a_z=0.0433$~\protect\cite{remarkerr2};
dotted lines show $(n_s/n)_{\hat{z}}$
for $a=0$ calculated in the grandcanonical ensemble.
Solid lines show the superfluid fraction given by Eq.~(\protect\ref{eq4}).
Dash-dotted lines show the condensate fraction $N_0/N$, 
and dot-dot-dot-dashed lines the fraction $N_{1D}/N$.
The inset of panel (a) shows 
$(n_s/n)_{\hat{x}}$ for $N=27$ and $T/T_c=0.2$ for three different
interaction strengths (using the same symbols as in the main
figure) as a function of $L$ on a logarithmic scale.
}
\label{fig3}
\end{figure}

We now turn to the calculation of the superfluid fraction 
$(n_s/n)_{\hat{n}}$ with 
respect to the axis $\hat{n}$.
For the non-interacting gas, the superfluid fraction with respect to, e.g., the $z$-axis, can be calculated
from the thermal expectation values of $x^2$
and $y^2$~\cite{stri96}.
To this aim, we consider a trapping geometry with
$\omega_y = \omega_x + \Delta \omega$ in the limit
$\Delta \omega \rightarrow 0$ (see Eq.~(7) of Ref.~\cite{stri96}).
Dotted lines in Fig.~\ref{fig3} show the resulting superfluid
fraction $(n_s/n)_{\hat{z}}$, calculated in the grandcanonical
ensemble, for (a) $L=1$,
(b) $L=10$ and (c) $L=100$ 
as a function of 
$T/T_c$ for $N=27$
non-interacting bosons.

Within the PIMC formulation, the superfluid fraction
$(n_s/n)_{\hat{n}}$ 
can be calculated from the square of the projected
area $A_{\hat{n}}$~\cite{sind89}, 
where $A_{\hat{n}}=\vec{A} \cdot \hat{n}$
and $\vec{A}$ denotes the area enclosed by the imaginary time 
paths~\cite{cepe95}.
Symbols in Fig.~\ref{fig3}
show the
superfluid fraction $(n_s/n)_{\hat{z}}$ 
calculated using the PIMC method for three different aspect ratios $L$;
diamonds show our 
results for $a=0$, squares those for $a=0.00433a_z$,
and triangles those for $a=0.0433a_z$.
For $a=0$,
the PIMC results for $(n_s/n)_{\hat{z}}$ (diamonds),
calculated in the canonical ensemble, agree well with
those calculated in the grandcanonical ensemble (dotted lines).

The superfluid fraction $(n_s/n)_{\hat{z}}$ 
is essentially one at small scaled temperatures, and
decreases gradually 
with increasing $T/T_c$. For $L=1$, $(n_s/n)_{\hat{z}}$
is about $0.05$ for $T/T_c=1$. For the larger aspect ratios
[see Figs.~\ref{fig3}(b) and (c)],
in contrast,
$(n_s/n)_{\hat{z}}$
is significantly larger at the
transition temperature (about $0.2$ for $L=10$ and about
0.65 for $L=100$). 
When plotted, as done here, as a function of 
$T/T_c$
the superfluid fraction 
$(n_s/n)_{\hat{z}}$
shows a very weak, if any, dependence on the 
interaction strength for all aspect ratios.

The spherically symmetric system with $L=1$
has no preferred symmetry axis, implying $(n_s/n)_{\hat{z}}=
(n_s/n)_{\hat{x}}$.
For $L>1$, however, the superfluid fraction $(n_s/n)_{\hat{x}}$
is distinctly different from $(n_s/n)_{\hat{z}}$.
Geometric arguments imply that
$(n_s/n)_{\hat{x}}$ approaches zero when the system
reaches the quasi-1D regime. When exposed to a rotation about $\hat{x}$,
the atoms move with the external trap, thus implying 
$\Theta^{rig}_{\hat{x}} = \Theta_{\hat{x}}$.
The inset of Fig.~\ref{fig3}(a)
shows the superfluid fraction
$(n_s/n)_{\hat{x}}$ for $N=27$ and $T/T_c=0.2$  for
three different scattering lengths (using the same symbols as
in the main figure) as a function of the aspect ratio $L$.
It is evident that $(n_s/n)_{\hat{x}}$
decreases rapidly with increasing $L$.

To connect with earlier work, we consider an analytical expression
for the superfluid fraction
$(n_s/n)_{\hat{z}}$, which
has been derived using the semi-classical approximation
for the non-interacting gas~\cite{stri96}. 
For a trap geometry with $\omega_y \approx \omega_x$ and
$\omega_z = \omega_x/L$, we generalize the treatment by Stringari~\cite{stri96}
to account for rotations about $\hat{n}=\hat{x}$.
To additionally improve the accuracy for small
$N$, we
use the $T_c$ that accounts for finite-size effects (see above),
\begin{eqnarray}
\label{eq4}
(n_s/n)_{\hat{n}}=
\frac{A \left[ 1- \left({T}/{T_c}\right)^3 \right] }
{1- \left(\frac{T}{T_c}\right)^3 + 
B \; 
\frac{1.80079}{L} \, \left(\frac{T}{T_c}\right)^4 \, 
\frac{k_B T_c}{\hbar \omega_z}}.
\end{eqnarray}
Here, $A$ and $B$ denote constants depending on the geometry
of the trap; $A=B=1$ for $\hat{n}=\hat{z}$,
and $A =1-[(1-L)/(1+L)]^2$ and 
$B=(L^2+1)/(L+1)$ for $\hat{n}=\hat{x}$.
Solid lines in Fig.~\ref{fig3} show the resulting approximate
superfluid fractions 
$(n_s/n)_{\hat{z}}$ (main figure)
and $(n_s/n)_{\hat{x}}$ [inset of Fig.~\ref{fig3}(a)].
The agreement 
between Eq.~(\ref{eq4}) and our
numerical results for $L=1$ is good. 
For $L=100$, however, Eq.~(\ref{eq4})
describes the superfluid fraction
only
qualitatively.
In particular, Eq.~(\ref{eq4}) clearly underestimates
$(n_s/n)_{\hat{z}}$ for
$T/T_c \gtrsim 1$.

For comparison, dash-dotted lines in Fig.~\ref{fig3} 
show the condensate fraction
$N_0/N$ calculated in the 
grandcanonical ensemble for the non-interacting gas.
For $L=1$, the condensate fraction roughly agrees with 
the superfluid fraction.
For $L=100$, in contrast, $N_0/N$ 
drops to zero at much lower temperatures than $(n_s/n)_{\hat{z}}$.
This shows that the condensate fraction and the superfluid fraction
are distinctly different quantities for highly-elongated systems.
For comparison, dot-dot-dot-dashed lines show the fraction of atoms $N_{1D}/N$
in the lowest transverse mode, where $N_{1D}= \sum_k N_{00k}$ and
$N_{ijk}/N$ denotes the fraction of atoms in the state
with $i$ quanta in the $x$-,
$j$ quanta in the $y$- and $k$ quanta in the $z$-direction.
For the highly-elongated gas with $L=100$,
the fraction of atoms $N_{1D}/N$ 
is larger than the superfluid fraction
$(n_s/n)_{\hat{z}}$ but shows a similar overall behavior.
Calculations for the non-interacting gas for $N=1000$
and $L=5000$ (not shown) indicate similar behaviors.

In sumary, this Letter describes microscopic calculations for
small Bose gases over a wide temperature range. 
We study the crossover from 3D to 1D 
by changing the trapping frequency. Since all our calculations 
are performed in full 3$N$-dimensional
configuration space, the freezing of the
radial motion at low temperatures emerges from our calculations;
it is not an input. 
Specifically,
we determine the temperature dependence of the quantum mechanical
moment of inertia.
This quantity has played a key role in the study of finite-size bosonic helium
droplets over the past 10 years or so~\cite{toen01}.
Measurements of the quantum mechanical moment of inertia of an impurity
embedded inside such a droplet have, e.g., 
shown unambigiously 
that bosonic
helium clusters with as few as about 60 atoms are superfluid~\cite{greb98}.
This paper shows that the effects of superfluidity are altered as the
effective dimensionality of the trapped gas changes from 3D to 1D. 
The superfluid fraction $(n_s/n)_{\hat{z}}$
is enhanced as the dimensionality is reduced. 
In the quasi-1D regime, the superfluid response is distinctly
different from the 
condensate fraction $N_0/N$ and very roughly follows 
the fraction $N_{1D}/N$ of atoms in the lowest transverse mode.

We gratefully acknowledge fruitful discussions with S. Giorgini 
and support by the NSF, grant PHY-0331529.


\begin{thebibliography}{10}

\bibitem{donn95}
R.~J. Donnelly,
\newblock Quantized Vortices in Helium II
\newblock (Cambridge University Press, Cambridge, 1995).

\bibitem{ande95}
M.~H. Anderson {\em{et al.}},
\newblock Science {\bf{269}},  198 (1995).

\bibitem{matt99}
M.~R. Matthews {\em{et al.}},
\newblock Phys. Rev. Lett. {\bf{83}},  2498 (1999).

\bibitem{abos01}
J.~R. Abo-Shaeer, C. Raman, J.~M. Vogels and W. Ketterle,
\newblock Science {\bf{292}},  476  (2001).

\bibitem{zwie05}
M. Zwierlein, Talk at workshop on ``Strongly Interacting Quantum Gases'';
April 18-21, 2005; Ohio Center for Theoretical Science.

%
\bibitem{baym69}
G. Baym,
\newblock in Mathematical Methods in Solid State and Superfluid Theory, edited
  by R. C. Clark and E. H. Derrick
\newblock (Oliver and Boyd, Edinburgh, 1969).

\bibitem{stri96}
S. Stringari,
\newblock Phys. Rev. Lett. {\bf{76}},   1405 (1996).

\bibitem{cepe95}
D.~M. Ceperley,
\newblock Rev. Mod. Phys. {\bf{67}},   279 (1995).

\bibitem{groo50}
S.~R. de Groot, G.~J. Hooyman and A. Seldam,
\newblock Proc. R. Soc. London, Ser. A {\bf{203}},   266 (1950).

\bibitem{sind89}
P. Sindzingre, M.~L. Klein and D.~M. Ceperley,
\newblock Phys. Rev. Lett. {\bf{63}},   1601 (1989).

\bibitem{bohr75}
A. Bohr and B.~R. Mottelson,
\newblock Nuclear Structure, Volume II, Nuclear Deformations
\newblock (W. A. Benjamin, Inc., Reading, 1975).

\bibitem{gerb04}
F. Gerbier {\em{et al.}},
\newblock Phys. Rev. Lett. {\bf{92}},  030405 (2004).

\bibitem{inou98}
S. Inouye {\em{et al.}},
\newblock Nature {\bf{392}},  151 (1998).

\bibitem{corn00}
S.~L. Cornish {\em{et al.}},
\newblock Phys. Rev. Lett. {\bf{85}}  1795 (2000).

\bibitem{goer01}
A. G\"orlitz {\em{et al.}},
\newblock Phys. Rev. Lett. {\bf{87}},  130402 (2001).

\bibitem{grei01}
M. Greiner {\em{et al.}},
\newblock Phys. Rev. Lett. {\bf{87}},  160405 (2001).

\bibitem{herz97}
C. Herzog and M. Olshanii,
\newblock Phys. Rev. A {\bf{55}},  3254 (1997).

\bibitem{dalf99}
F. Dalfovo, S. Giorgini, L.~P. Pitaevskii and S. Stringari,
\newblock Rev. Mod. Phys. {\bf{71}},  463 (1999).

\bibitem{petr04}
see, e.g., D.~S. Petrov, D.~M. Gangardt and G.~V. Shlyapnikov,
\newblock J. Phys. IV (France) {\bf{116}}, ~3 (2004).

\bibitem{drut97}
N.~J. {van Druten} and W. Ketterle,
\newblock Phys. Rev. Lett. {\bf{79}},  549 (1997).

\bibitem{kett96}
W. Ketterle and N.~J. {van Druten},
\newblock Phys. Rev. A {\bf{54}},  656 (1996).

\bibitem{remarkerr1}
{Expectation values calculated by the PIMC method have an errorbar since the
  technique is stochastic in nature.}

\bibitem{remarkerr2}
{The PIMC data for small $T/T_c$ have the largest errorbars, the largest of
  which is at most four times the symbol size. }

\bibitem{toen01}
J.~P. Toennies, A.~V. Vilesov and K.~B. Whaley,
\newblock Phys. Today {\bf{54}}, ~31 (2001).

\bibitem{greb98}
S. Grebenev, J.~P. Toennies and A.~F. Vilesov,
\newblock Science {\bf{279}},  2083 (1998).

\end{thebibliography}
\end{document}